\documentstyle[tighten,preprint,aps,prb]{revtex} 

\begin{document}
\draft
\title{Effects of uniaxial strain in ${\rm {\bf LaMnO_3}}$}
\author{K. H. Ahn \cite{Ahn} and A. J. Millis}
\address{Center for Materials Theory\\
Department of Physics and Astronomy, Rutgers University\\
Piscataway, New Jersey 08854}
\maketitle

\begin{abstract}
The effects of uniaxial strain on the structural, orbital, optical, and
magnetic properties of ${\rm LaMnO_{3}}$ are calculated using a general
elastic energy expression, along with a tight-binding parameterization of
the band theory. Tensile uniaxial strain of the order of
2 \% (i.e., of the
order of magnitude of those induced in thin films by lattice mismatch with
substrates) is found to lead to changes in the magnetic ground state,
leading to dramatic changes in the band structure and optical conductivity
spectrum. The magnetostriction effect associated with the Neel 
transition of
bulk(unstrained) ${\rm LaMnO_{3}}$ is also determined. Due to the
Jahn-Teller coupling, the uniform tetragonal distortion mode is softer 
in $%
{\rm LaMnO_{3}}$ than in doped cubic manganates. Reasons why the 
observed ($%
\pi \pi 0$) orbital ordering is favored over a ($\pi \pi \pi $) 
periodicity
are discussed.
\end{abstract}


\narrowtext

\newpage

\section{Introduction}

The `colossal' magnetoresistive manganese perovskites have been a focus of
recent attention.\cite{Royal} Since most of the technological applications
require thin films on substrates, it is important to understand the 
effects
of strains induced by substrates. Because Mn $e_{g}$ electrons, which
determine important physical properties of these materials, are coupled to
the lattice degrees of freedom through the Jahn-Teller (JT) coupling, 
it is
expected that uniaxial or biaxial strains are important and that even
relatively small strains may result in observable effects on the 
electronic
properties of these materials. Recently, the effects of substrate-induced
strains on the properties of thin manganate films have been studied
experimentally.\cite{Millis.strain,strain} It is indeed found that the
ferromagnetic and metal-insulator transition temperature, $T_{c}$, depends
sensitively on the volume-preserving uniaxial strains, as do the magnetic
anisotropy, magnetoresistance, and charge ordering transition.

In this work, we study the effects of uniaxial strains in 
${\rm LaMnO_{3}}$,
which is the parent compound of the doped manganese perovskites. Our
motivations are to further clarify the properties of this interesting
compound and to test techniques and fix parameters so the more complicated
behaviors of the doped compounds may be studied. At very high 
temperatures,
bulk ${\rm LaMnO_{3}}$ exists in cubic perovskite structure, but at T $<$
750 K, it has a static $(\pi ,\pi ,0)$ $3x^{2}-r^{2}/3y^{2}-r^{2}$ type
Jahn-Teller distortion.\cite{Ellemans71} It also has a uniform tetragonal
distortion, which originates from the coupling of the staggered and 
uniform
distortions by an anharmonic elastic energy. \cite{Kanamori60} This 
coupling
implies that the substrate-induced strain may affect the orbital ordering.

In Ref. \onlinecite{Millis96}, following Kanamori, a model with harmonic
Mn-O and Mn-Mn elastic forces and a local anharmonic energy term was 
used to
study lattice distortions in this material. It predicted $(\pi ,\pi )$ 
type
ordering in $xy$ plane as observed in ${\rm LaMnO_{3}}$. However, 
according
to this model, $(\pi ,\pi ,0)$ and $(\pi ,\pi ,\pi )$ type orderings have
the same energy. The reason why $(\pi ,\pi ,0)$ ordering is favored 
has not
been understood so far. Bulk ${\rm LaMnO_{3}}$ has so called A-type
antiferromagnetic (AF) ordering below 140 K, in which spins align parallel
in $xy$ plane and antiparallel along $z$ direction.\cite{Ellemans71} This
peculiar spin ordering pattern is the result of the exchange interaction
between Mn ions which depends on the $e_{g}$ orbital ordering as explained
in Ref. \onlinecite{Millis97}. Indications of a coupling between orbital
ordering and magnetic ordering have also been found in a recent X-ray
resonant scattering experiment.\cite{Murakami98}

In this paper, we present more general expressions for the elastic energy
than those in Ref. \onlinecite{Millis96}. We use these to study the ground
state energy and distortions in bulk state, and the effects of uniaxial
strains in thin films. We also present a simple model of the magnetic
interaction depending explicitly on the orbital states, and we use this to
study the magnetostriction effects in bulk state and the change of the
magnetic interaction due to the strains in thin films. We examine the
changes in the band structure and optical conductivity due to the strains,
using a nearest-neighbor tight binding approximation. We also compare the
energies of $(\pi ,\pi ,0)$ and $(\pi ,\pi ,\pi )$ type orderings, and
examine why the observed $(\pi ,\pi ,0)$ distortion is favored over the $%
(\pi ,\pi ,\pi )$ type distortion. Our calculations suggest that 2 \%
tensile strain can change the observed A-type(layered) antiferromagnetic
state into a purely antiferromagnetic state. This change would induce 
large
changes in band structure and optical conductivity spectrum, which we
calculate. Finally, we show that the magnetostriction effect at the Neel
transition is large.

The rest of the paper is organized in the following way. Section II
introduces a model of elastic energy, Sect. III a model of the magnetic
interaction and magnetostriction effect, and Sect. IV a tight binding 
model
for band structure and optical conductivity. Section V compares the ground
state energies of $(\pi ,\pi ,0)$ and $(\pi ,\pi ,\pi )$ type distortions.
Section VI presents the results. Section VII summarizes the main
conclusions. In Appendices A and B, we show how we determine general
expressions of the elastic energy, and how we determine the parameters,
respectively.

\section{Model of lattice energy}

\subsection{Overview}

To study the effects of uniaxial strains in thin films, we first need to
understand the strains already present in bulk state. The elastic energy
depends on three important variables: the oxygen displacement along Mn-Mn
direction, the three-dimensional Mn ion displacement, and the Mn $e_{g}$
electron orbital state.\cite{Kanamori60}$^{\text{,}}$\cite{Millis96} More
precisely, the degrees of freedom we will consider are defined in the
following way: In the ideal cubic perovskite structure with lattice 
constant 
$a_{0}$, the Mn ions are located at $a_{0}\vec{i}$, and oxygen ions at $%
a_{0}(\vec{i}+\hat{a}/2)$, where $i_{x}$, $i_{y}$, and $i_{z}$ are 
integers
and $\hat{a}$=$\hat{x}$, $\hat{y}$, and $\hat{z}$. We write the 
displacement
of Mn at $a_{0}\vec{i}$ as 
$a_{0}(\vec{e}\cdot \vec{i}+\vec{\delta}_{\vec{i}%
})$, where $\vec{\delta}_{\vec{i}}$ represents nonzero-wavevector Mn-ion
displacement, $\vec{e}=e_{xx}\hat{x}+e_{yy}\hat{y}+e_{zz}\hat{z}$, and $%
e_{ij}$ is the conventional strain tensor referred to the ideal cubic
perovskite lattice (we need only the diagonal components). The 
displacement
of oxygen at $a_{0}(\vec{i}+\hat{a}/2)$ along Mn-Mn axis is 
$a_{0}[\vec{e}%
\cdot (\vec{i}+\hat{a}/2)+u_{\vec{i}}^{a}\hat{a}]$, where 
$u_{\vec{i}}^{a}$
represents O-ion displacement with nonzero wavevector. Figure 1 
shows $\vec{%
\delta}_{\vec{i}}$ and $u_{\vec{i}}^{x,y,z}$. We assume that we have 
already
minimized the elastic energy over the displacements of O ions 
perpendicular
to the Mn-Mn axis and the displacement of La ions. Therefore, even 
though we
do not consider the buckling of the Mn-O-Mn bond explicitly, its effect 
is
implicitly included in our harmonic and anharmonic elastic 
constants below.

We treat the elastic energy due to these strains in up to cubic anharmonic
terms, and the coupling of the strain to the Mn $e_{g}$ orbital state 
by the
Jahn-Teller coupling. In this paper, instead of representing the elastic
energy in terms of spring constants between different ions as done 
in Refs. %
\onlinecite{Kanamori60} and \onlinecite{Millis96}, we will introduce a 
more
general and, we hope, more useful formulation in which long wavelength
lattice distortions are treated via conventional elastic theory, while the
short wavelength atomic motions are treated explicitly.

\subsection{General energy expressions for $(\pi ,\pi ,0)$ and 
$(\pi ,\pi
,\pi )$ distortions}

The energy per Mn ion due to uniform strains, $e_{xx}$, $e_{yy}$, and $%
e_{zz} $, is most conveniently written in terms of the following
combinations: 
\begin{eqnarray}
Q_{1u} &=&\frac{a_{0}}{\sqrt{3}}(e_{xx}+e_{yy}+e_{zz}), \\
Q_{2u} &=&\frac{a_{0}}{\sqrt{2}}(e_{xx}-e_{yy}), \\
Q_{3u} &=&\frac{a_{0}}{\sqrt{6}}(2e_{zz}-e_{xx}-e_{yy}).
\end{eqnarray}
We have 
\begin{equation}
\frac{E_{u}}{N_{Mn}}=\frac{1}{2}K_{B}Q_{1u}^{2}+\frac{1}{2}%
K^{*}(Q_{2u}^{2}+Q_{3u}^{2}),
\end{equation}
where the bulk modulus $K_{B}=a_{0}(c_{11}+2c_{12})=3a_{0}c_{B}$, the
Jahn-Teller shear modulus $K^{*}=a_{0}(c_{11}-c_{12})=2a_{0}c^{*}$, and $%
c_{ij}$ are the usual elastic constants.

In addition to these uniform strains, we consider staggered distortions 
with
wave vector $\vec{k}=(\pi ,\pi ,0)$ or $(\pi ,\pi ,\pi ).$ We represent 
the
amplitudes of the distortions by $a_{0}\vec{\delta}_{\vec{k}}$ for Mn ions
and $a_{0}u_{\vec{k}}^{x,y,z}$ for O ions. Translational symmetry implies
that the uniform distortion and the staggered distortion do not couple 
with
each other up to the second order. Therefore, the harmonic elastic energy
due to the staggered distortions simply adds to the harmonic uniform 
strain
energy. In Appendix A, we use symmetry arguments to obtain the general 
forms
of the elastic energies due to $\vec{\delta}_{\vec{k}}$ and $u_{\vec{k}%
}^{x,y,z}$ for $(\pi ,\pi ,0)$ and $(\pi ,\pi ,\pi )$ distortions. In the
Jahn-Teller coupling energy $E_{JT}$, which we will introduce later, only
the lattice distortions which have even parity about Mn ions appear.
Therefore, $E_{JT}$ does not depend on the Mn-distortions 
$\delta _{\vec{k}%
}^{x,y,z}$ for either $\vec{k}=(\pi ,\pi ,0)$ or 
$\vec{k}=(\pi ,\pi ,\pi )$,
or the z-direction oxygen distortion $u_{\vec{k}}^{z}$ for $\vec{k}=(\pi
,\pi ,0)$. The energy cost of the relevant distortions is most 
conveniently
written in terms of 
\begin{eqnarray}
Q_{2s} &=&\frac{a_{0}}{\sqrt{2}}(v_{sx}-v_{sy}), \\
Q_{3s} &=&\frac{a_{0}}{\sqrt{6}}(2v_{sz}-v_{sx}-v_{sy}),
\end{eqnarray}
where $v_{sa}$ is the $(\pi \pi 0)$ or $(\pi \pi \pi )$ amplitude of $v_{%
\vec{i}}^{a}$ $=u_{\vec{i}}^{a}-u_{\vec{i}-\hat{a}}^{a}$. The energy of a
staggered distortion, $E_{s}$, depends upon the ordering wavevector 
and, for
the distortions we consider, is 
\begin{eqnarray}
\frac{E_{s}(\pi ,\pi ,0)}{N_{Mn}} &=&\frac{1}{2}K_{2s}Q_{2s}^{2}+
\frac{1}{2}%
K_{3s}Q_{3s}^{2}, \\
\frac{E_{s}(\pi ,\pi ,\pi )}{N_{Mn}} &=&\frac{1}{2}K_{s}\left(
Q_{2s}^{2}+Q_{3s}^{2}\right) ,
\end{eqnarray}
where the $K_{2s\text{, }}K_{3s\text{, }}$and $K_{s}$ are the elastic
constants defined in Appendix A, and arise mainly from the Mn-O bond
stretching mode. \cite{Es}

We compare the sizes of $K^{*}$, $K_{2s}$, $K_{3s}$, and $K_{s}$. This
comparison is important to determine the ground state distortions. 
A uniform
strain changes more bonds than a staggered distortion. For example, 
within a
harmonic nearest-neighbor approximation, $Q_{2s}$ and $Q_{3s}$ modes 
involve
only the Mn-O bond, but $Q_{2u}$ and $Q_{3u}$ modes involve both Mn-O and
Mn-Mn bonds. Therefore, uniform modes have larger elastic moduli than
staggered modes, which remains true when reasonable further neighbor
interactions are included. Therefore, we expect $K^{*}>K_{3s},$ 
$K_{2s}$, $%
K_{s}$. Our analysis of the general expression of the elastic energy given
in Appendix A shows $K_{3s}$ $>K_{2s}$.

The staggered lattice distortion is caused by the Jahn-Teller coupling to
the Mn $e_{g}$ orbital state. The Mn $e_{g}$ electron state on site 
$a_{0}%
\vec{i}$ is represented by 
\begin{equation}
|\theta _{\vec{i}}>=\cos \theta _{\vec{i}}|3z^{2}-r^{2}>+\sin 
\theta _{\vec{i%
}}|x^{2}-y^{2}>,
\end{equation}
where $|3z^{2}-r^{2}>$ and $|x^{2}-y^{2}>$ are the two linearly 
independent $%
e_{g}$ orbitals on the site. The cases of interest here are two-sublattice
distortions. We represent the orbital states of $e_{g}$ electrons on these
two sublattices as $\theta _{1}$ and $\theta _{2}$. Then the JT energy is%
\cite{Kanamori60}$^{\text{,}}$\cite{Millis96} 
\begin{equation}
\frac{E_{JT}}{N_{Mn}}=-\sqrt{\frac{3}{2}}\lambda \frac{1}{2}\left[ \cos
2\theta _{1}(Q_{3u}+Q_{3s})+\sin 2\theta _{1}(Q_{2u}+Q_{2s})+\cos 2\theta
_{2}(Q_{3u}-Q_{3s})+\sin 2\theta _{2}(Q_{2u}-Q_{2s})\right] .
\end{equation}
To represent the energy only in terms of lattice distortions, we minimize
the above Jahn-Teller energy with respect to the orbital states 
$\theta _{1}$
and $\theta _{2}$. We obtain 
\begin{eqnarray}
\cos 2\theta _{1,2}^{{\rm min}} &=&
\frac{Q_{3u}\pm Q_{3s}}{\sqrt{(Q_{2u}\pm
Q_{2s})^{2}+(Q_{3u}\pm Q_{3s})^{2}}}, \\
\sin 2\theta _{1,2}^{{\rm min}} &=&
\frac{Q_{2u}\pm Q_{2s}}{\sqrt{(Q_{2u}\pm
Q_{2s})^{2}+(Q_{3u}\pm Q_{3s})^{2}}},
\end{eqnarray}
and 
\begin{equation}
\frac{E_{JT}}{N_{Mn}}=-\frac{1}{2}\sqrt{\frac{3}{2}}\lambda \left[ \sqrt{%
(Q_{2u}+Q_{2s})^{2}+(Q_{3u}+Q_{3s})^{2}}+\sqrt{%
(Q_{2u}-Q_{2s})^{2}+(Q_{3u}-Q_{3s})^{2}}\right] .
\end{equation}

We also include the largest anharmonic energy, which is the one between 
the
nearest neighbor Mn-O pair. It is given by 
\begin{equation}
E_{anh}=\frac{4}{\sqrt{3}}Aa_{0}^{3}\sum_{i,a}\left( \frac{e_{aa}}{2}%
+u_{i}^{a}-\delta _{i}^{a}\right) ^{3}+
\left( \delta _{i}^{a}-u_{i-a}^{a}-%
\frac{e_{aa}}{2}\right) ^{3}.  \label{Eanh}
\end{equation}
Total energy is the sum of the terms considered so far: 
\begin{equation}
E_{tot}^{elastic}=E_{u}+E_{s}+E_{JT}+E_{anh}.
\end{equation}
By minimizing $E_{tot}^{elastic}$, we find the distortions induced by 
the JT
coupling and anharmonic energy terms, which will be discussed below for $%
\vec{k}=(\pi ,\pi ,0)$ and $\vec{k}=(\pi ,\pi ,\pi )$ distortions 
separately.

\subsection{Energy minimization for $(\pi ,\pi ,0)$ distortion}

We minimize $E_{u}+E_{s}+E_{JT}$ and then treat $E_{anh}$ as a 
perturbation,
since we expect and will show below that the anharmonic term is small
compared to the harmonic terms. We find that in the ground state of $%
E_{u}+E_{s}+E_{JT}$, the distortion mode which has the smallest modulus
among $Q_{2u}$, $Q_{3u}$, $Q_{2s}$, and $Q_{3s}$ is non-zero, and all the
other distortion modes vanish. Since, we found $K^{*}>K_{3s}>K_{2s},$ the
ground state of $E_{u}+E_{s}+E_{JT}$ is 
\begin{eqnarray}
Q_{3s}=Q_{2u}=Q_{3u} &=&0, \\
Q_{2s} &=&\sqrt{\frac{3}{2}}\frac{\lambda }{K_{2s}}, \\
E_{{\rm min}} &=&-\frac{3}{4}\frac{\lambda ^{2}}{K_{2s}}, \\
{\theta _{1},\theta _{2}} &=&{\ \pi /4,3\pi /4}.
\end{eqnarray}
It is noteworthy that in this ground state the Mn lattice itself preserves
cubic symmetry, and only oxygen ions make staggered distortions.

We next 
study how the anharmonic energy term changes the above ground state.
We represent $E_{anh}$ in Eq. (\ref{Eanh}) by $\vec{\delta}_{\vec{k}}$, 
$u_{%
\vec{k}}^{x,y,z}$, and $e_{xx,yy,zz}$. Direct expansion (or symmetry
argument) shows that each of $\vec{\delta}_{\vec{k}}$ and 
$u_{\vec{k}}^{z}$
appears only as the second order in $E_{anh}$, which implies that these
distortions remain zero unless the coupled uniform strains exceed certain
values and the lattice becomes unstable. After representing 
$E_{anh}(\pi \pi
0)$ in terms of $Q_{1u,2u,3u,2s,3s}$, the same argument shows $%
Q_{3s}=Q_{2u}=0$. By taking the largest remaining term, we obtain\cite
{fn:Q2u}

\begin{equation}
\frac{E_{anh}(\pi \pi 0)}{N_{Mn}}\approx 
AQ_{2s}^{2}(Q_{1u}-\frac{1}{\sqrt{2}%
}Q_{3u}).
\end{equation}

The total energy for $\vec{k}=(\pi ,\pi ,0)$ distortion which we will
minimize to find strains in bulk state or in thin films is 
\begin{eqnarray}
E_{{\rm tot}}(\pi \pi 0) &=&(E_{u}+E_{s}+E_{JT}+E_{anh})/N_{\text{Mn}} 
\nonumber \\
&=&\frac{1}{2}K_{B}Q_{1u}^{2}+\frac{1}{2}K^{*}Q_{3u}^{2}+\frac{1}{2}%
K_{2s}Q_{2s}^{2}-\sqrt{\frac{3}{2}}\lambda \sqrt{Q_{2s}^{2}+Q_{3u}^{2}}%
+AQ_{2s}^{2}\left( Q_{1u}-\frac{1}{\sqrt{2}}Q_{3u}\right) .  \label{Etot}
\end{eqnarray}
In bulk state, there is no external constraint. When we minimize 
$E_{{\rm tot%
}}(\pi \pi 0)$ in a leading order in $A$ to find the lattice distortions 
in
the bulk state, we obtain 
\begin{equation}
E_{tot}^{MIN}(\pi \pi 0)=-\frac{3}{4}\frac{\lambda ^{2}}{K_{2s}}-
A^{2}\frac{9%
}{16}\frac{\lambda ^{4}}{K_{2s}^{4}}\frac{2K^{*}-2K_{2s}+K_{B}}{%
K_{B}(K^{*}-K_{2s})}+O(A^{4}),  \label{eq:Eminpipi0}
\end{equation}
for 
\begin{eqnarray}
Q_{1u} &=&-\frac{3}{2}\frac{\lambda ^{2}}{K_{2s}}\frac{A}{K_{2s}^{2}}%
+O(A^{3}), \\
Q_{2s} &=&\sqrt{\frac{3}{2}}\frac{\lambda }{K_{2s}}+O(A^{2}), \\
Q_{3s} &=&Q_{2u}=0, \\
Q_{3u} &=&
\frac{3\lambda ^{2}A}{2\sqrt{2}(K^{*}-K_{2s})K_{2s}^{2}}+O(A^{3}).
\label{eq:Q3upipi0}
\end{eqnarray}
The results show that the observed uniform tetragonal distortion is due to
the anharmonic term which couples staggered and uniform distortions. 
${\rm %
LaMnO_{3}}$ expands upon heating, which implies $A<0$. Therefore, above
result indicates $Q_{3u}<0$, which is consistent with the observed
distortion in bulk ${\rm LaMnO_{3}}$. Since $K^{*}-K_{2s}$ is order of
magnitude smaller than $K_{2s}$, $Q_{3u}$ is order of magnitude larger 
than $%
Q_{1u}$. Up to order of $A$, the anharmonic term does not change $Q_{2s}$.

We note ``$Q_{3u}$ mode softening'' : When we have lattice distortion $%
Q_{2s}=\sqrt{3/2}\frac{\lambda }{K_{2s}}$, the JT coupling term 
in Eq. (\ref
{Etot}), expanded about small $Q_{3u}$, effectively reduces $Q_{3u}$ mode
modulus by $-K_{2s}$. Therefore, when the anharmonic term induces 
$Q_{3u}$,
the restoring spring constant is $K^{*}-K_{2s}$ rather than $K^{*}$. 
This is
the reason why we have $K^{*}-K_{2s}$ in the denominator of Eq. (\ref
{eq:Q3upipi0}) and have a relatively large $Q_{3u}$($\sim $ 35 \% of 
$Q_{2s}$
from crystallography data). Thus we expect the shear modulus corresponding
to the $Q_{3u}$ distortion to be much smaller in ${\rm LaMnO_{3}}$ than in
doped cubic manganates.

\subsection{Energy minimization for $(\pi ,\pi ,\pi )$ distortion}

By applying similar considerations to $(\pi ,\pi ,\pi )$ distortion, and
using the condition $K^{*}>K_{s}$, we find that the degenerate ground 
states
of $E_{u}+E_{s}(\pi \pi \pi )+E_{JT}(\pi \pi \pi )$ are 
\begin{eqnarray}
Q_{2u}=Q_{3u} &=&0, \\
Q_{2s} &=&\sqrt{\frac{3}{2}}\frac{\lambda }{K_{s}}\sin 2\theta _{1}, \\
Q_{3s} &=&\sqrt{\frac{3}{2}}\frac{\lambda }{K_{s}}\cos 2\theta _{1}, \\
E_{{\rm min}} &=&-\frac{3}{4}\frac{\lambda ^{2}}{K_{s}}, \\
\theta _{2} &=&\pi -\theta _{1},
\end{eqnarray}
where $\theta _{1}$ is an arbitrary angle between 0 and $\pi $. After we
include the same anharmonic energy and apply the same arguments used for 
the 
$(\pi ,\pi ,0)$ distortion, we find the total energy expression for the $%
(\pi ,\pi ,\pi )$ distortion which we will minimize further is 
\begin{eqnarray}
E_{tot}(\pi \pi \pi ) &=&[E_{u}+E_{s}(\pi \pi \pi )+E_{JT}(\pi \pi \pi
)+E_{anh}(\pi \pi \pi )]/N_{Mn} \\
&=&\frac{1}{2}K_{B}Q_{1u}^{2}+\frac{1}{2}K^{*}(Q_{2u}^{2}+Q_{3u}^{2})+
\frac{1%
}{2}K_{s}(Q_{2s}^{2}+Q_{3s}^{2}) \\
&&-\sqrt{\frac{3}{2}}\lambda \frac{1}{2}\left[ \sqrt{%
(Q_{2u}+Q_{2s})^{2}+(Q_{3u}+Q_{3s})^{2}}+\sqrt{%
(Q_{2u}-Q_{2s})^{2}+(Q_{3u}-Q_{3s})^{2}}\right]  \nonumber \\
&&+A\left[ Q_{2s}^{2}(Q_{1u}-\frac{1}{\sqrt{2}}Q_{3u})+Q_{3s}^{2}(Q_{1u}+%
\frac{1}{\sqrt{2}}Q_{3u})-\sqrt{2}Q_{2s}Q_{3s}Q_{2u}\right] .  \nonumber
\end{eqnarray}
When we minimize $E_{tot}(\pi \pi \pi )$, we obtain 
\begin{equation}
E_{tot}^{MIN}(\pi \pi \pi )=-\frac{3}{4}\frac{\lambda ^{2}}{K_{s}}-A^{2}%
\frac{9}{16}\frac{\lambda ^{4}}{K_{s}^{4}}\frac{2K^{*}-2K_{s}+K_{B}}{%
K_{B}(K^{*}-K_{s})}+O(A^{4})  \label{eq:Eminpipipi}
\end{equation}
for 
\begin{eqnarray}
Q_{1u} &=&-\frac{3}{2}\frac{\lambda ^{2}}{K_{s}}\frac{A}{K_{s}^{2}}+
O(A^{3}),
\\
Q_{2s} &=&\sqrt{\frac{3}{2}}\frac{\lambda }{K_{s}}+O(A^{2}), \\
Q_{3s} &=&Q_{2u}=0, \\
Q_{3u} &=&\frac{3\lambda ^{2}A}{2\sqrt{2}(K^{*}-K_{s})K_{s}^{2}}+O(A^{3}).
\end{eqnarray}
The other two physically equivalent distortions obtained by permuting 
$x$, $%
y $, and $z$ from above results are also degenerate ground states. The
ground state energies for the ($\pi \pi 0$) and ($\pi \pi \pi $) 
distortions
will be compared in Sect. V.

\subsection{Energy minimization for strained films}

In this paper, by uniaxial strain, we mean a tetragonal strain with axis
perpendicular to film plane. This strain can be applied by growing 
epitaxial
films on square lattice substrates with lattice parameters different from
the $xy$ plane lattice parameter for bulk ${\rm LaMnO_{3}}$. For thin 
films
with uniaxial strains, we calculate the lattice distortions in the 
following
way: We assume that ($\pi \pi 0$) distortion pattern is favored even in
strained films. We also assume perfect epitaxy; therefore $e_{xx}$ and $%
e_{yy}$ are determined by substrates. Then $Q_{2s}$ and $e_{zz}$ for given
strains can be found by minimizing $E_{tot}$ about these distortions, 
which
also give $Q_{1u}$ and $Q_{3u}$. From $Q_{2s}$ and $Q_{3u}$, we find 
the $%
e_{g}$ orbital states, $\theta _{1}$ and $\theta _{2}$, by 
\begin{eqnarray}
\cos (2\theta _{1}) &=&\frac{Q_{3u}}{\sqrt{Q_{2s}^{2}+Q_{3u}^{2}}},
\label{eq:th} \\
\theta _{2} &=&\pi -\theta _{1}.
\end{eqnarray}
Since the external strain changes $Q_{2s}$ and $Q_{3u}$, it also changes 
the
JT splitting, given by $\delta \Delta E_{JT}=\sqrt{6}\lambda \sqrt{%
Q_{2s}^{2}+Q_{3u}^{2}}-\sqrt{6}\lambda \sqrt{%
(Q_{2s}^{eq})^{2}+(Q_{3u}^{eq})^{2}}.$ Parameters of the model, i.e., 
$K_{B}$%
, $K^{*}$, $K_{2s}$, $\lambda $, and $A,$ are determined from experiments,
as explained in Appendix B. The determined parameter values are shown in
Table I.

\section{Model of magnetic interaction and magnetostriction effects}

\subsection{Magnetic interaction}

In this section, we present a model describing magnetism in bulk and
strained films of LaMnO$_{3}$. The superexchange interaction between Mn 
ions
in ${\rm LaMnO_{3}}$ depends on the electron orbital overlap, particularly
the overlap between Mn $e_{g}$ orbitals and O $p$ orbitals.\cite
{Mryasov97.Solovyev96} It is also argued that $t_{2g}$ electrons also
contribute antiferromagnetic interaction. We build a simple model below,
which incorporates these two contributions of exchange interactions.

We calculate the superexchange interaction due to the $e_{g}$ electrons
using a similar method as in Ref. \onlinecite{Millis97}. We assume $e_{g}$
spin is always parallel to $t_{2g}$ spin at each site due to the strong
Hund's coupling. For the two Mn ions, one at $\vec{i}$ and the other at $%
\vec{i}+\hat{z}$, we find 
\begin{equation}
J_{e_{g}}(\theta _{1},\theta _{2})=-J_{F}\left( \sin ^{2}\theta _{1}\cos
^{2}\theta _{2}+\cos ^{2}\theta _{1}\sin ^{2}\theta _{2}\right) ,
\label{eq:Jeg}
\end{equation}
where $J_{F}$ is a positive parameter of the model. This is, in fact,
equivalent to the model in Ref. \onlinecite{Millis97}, if we assume that 
the
state with two holes on the intermediate oxygen ion (which was 
considered in
Ref. \onlinecite{Millis97}) requires an infinite energy. Equation (\ref
{eq:Jeg}) shows that when one of $\theta _{1}$ and $\theta _{2}$ is zero
(i.e., $|3z^{2}-r^{2}>$ state) and the other is $\pi /2$ (i.e., $%
|x^{2}-y^{2}>$ state), $J_{e_{g}}(\theta _{1},\theta _{2})$ is most
ferromagnetic due to the maximized hopping between the filled orbital 
on one
site and the empty orbital on the other site. When $\theta _{1}=\theta
_{2}=\pi /2$, or $\theta _{1}=\theta _{2}=0$, $J_{e_{g}}(\theta _{1},
\theta
_{2})$ is least ferromagnetic, since the hopping between the filled and
empty orbitals on the two sites vanishes. The $t_{2g}$ superexchange is
expected to be independent of $\theta _{1}$ and $\theta _{2}$, and always
antiferromagnetic, so we set $J_{t_{2g}}=J_{AF}$. Therefore, along the $z$
direction the total exchange interaction is 
\begin{eqnarray}
J_{z} &=&J_{t_{2g}}+J_{e_{g}}  \nonumber \\
&=&J_{AF}-J_{F}\left( \sin ^{2}\theta _{1}\cos ^{2}\theta _{2}+\cos
^{2}\theta _{1}\sin ^{2}\theta _{2}\right) .
\end{eqnarray}
The sign of the total superexchange is determined by the competition 
between
the $e_{g}$ ferromagnetism and $t_{2g}$ antiferromagnetism. Along $x$ 
and $y$
directions, proper rotations result in 
\begin{eqnarray}
J_{x} &=&J_{AF}-J_{F}\left( \sin ^{2}(\theta _{1}-\frac{2\pi }{3})\cos
^{2}(\theta _{2}-\frac{2\pi }{3})+\cos ^{2}(\theta _{1}-
\frac{2\pi }{3})\sin
^{2}(\theta _{2}-\frac{2\pi }{3})\right) , \\
J_{y} &=&J_{AF}-J_{F}\left( \sin ^{2}(\theta _{1}+\frac{2\pi }{3})\cos
^{2}(\theta _{2}+\frac{2\pi }{3})+
\cos ^{2}(\theta _{1}+\frac{2\pi }{3})\sin
^{2}(\theta _{2}+\frac{2\pi }{3})\right) .
\end{eqnarray}
Using the condition $\theta _{2}=\pi -\theta _{1}$ obtained before, we 
find 
\begin{eqnarray}
J_{x} &=&J_{y}\equiv J_{xy}=J_{xy}^{0}+J^{*}\cos 4\theta _{1}, \\
J_{z} &=&J_{z}^{0}+J^{*}\cos 4\theta _{1},
\end{eqnarray}
where $J_{xy}^{0}=J_{AF}-5J_{F}/8$, $J_{z}^{0}=J_{AF}-J_{F}/4$, and $%
J^{*}=J_{F}/4$. It explicitly shows how the magnetic coupling depends 
on the
orbital states.

\subsection{Magnetostriction effects}

In this section, we present our model of magnetostriction effect in bulk
LaMnO$_{3}$. We expect magnetostriction effect for the following reason:
Since $J_{x}$, $J_{y}$, and $J_{z}$ depend on the orbital states $\theta
_{1} $ and $\theta _{2}$, once certain magnetic ordering occurs, the
magnetic ordering, in turn, will change $\theta _{1}$ and $\theta _{2}$ to
gain further magnetic energy. Through the JT coupling, this change in the
orbital states can cause the change in the JT strains. Magnetism and 
lattice
distortion are coupled through the Mn $e_{g}$ orbital degree of freedom.

To estimate the size of the magnetostriction effects, we add to Eq. (\ref
{Etot}) the term 
\begin{equation}
\frac{E_{mag}}{N_{Mn}}=-4\left( 2|J_{xy}(\theta _{1})|+|J_{z}(\theta
_{1})|\right) ,
\end{equation}
which represents the mean-field T=0 magnetic energy, and find the 
changes in
the orbital states. For A-type antiferromagnetic state, after dropping
constant terms, we obtain 
\begin{equation}
\frac{E_{mag}}{N_{Mn}}=4J^{*}\cos 4\theta _{1}.
\end{equation}
By adding $E_{mag}/N_{Mn}$ to the total energy, we can find the extra
structure at T=0 due to the magnetic order.

Using the Landau free energy method, we examine whether the 
magnetostriction
effect makes the phase transition first order or not. From the energy gain
due to the magnetic order at T=0, and the mean field estimate 
of $T_{c}$, we
obtain the Landau free energy per site, $f$ : 
\begin{equation}
f(\theta _{1},Q_{1u},Q_{2s},Q_{3u},m)=2[T-T_{c}(\theta
_{1})]m^{2}+T_{c}(\theta _{1})m^{4}+E_{elastic}(\theta
_{1},Q_{1u},Q_{2s},Q_{3u}),  \label{eq:Landauf}
\end{equation}
where $m=M/M(T=0)$ is the normalized magnetization and $E_{elastic}(\theta
_{1},Q_{1u},Q_{2s},Q_{3u})$ is the total elastic energy obtained in Sect.
II. Magnetic ordering temperature, $T_{c}(\theta _{1})$, can be obtained 
by
molecular field theory:\cite{aschcroft} 
\begin{equation}
T_{c}(\theta _{1})=2\left( 4|J_{xy}(\theta _{1})|+2|J_{z}(\theta
_{1})|\right) .
\end{equation}

We can find the order of the phase transition and the size of the
magnetostriction effect in the following approximation: Since $Q_{1u}$ is
not directly coupled to $\theta _{1}$, we expect the change in $Q_{1u}$ is
small, which can be seen from the numerical results in Table II. 
Therefore,
we neglect $Q_{1u}$ dependence and obtain $E_{elastic}(\theta
_{1},Q_{2s},Q_{3u})$, which we minimize further about $Q_{2s}$ and 
$Q_{3u}$
to obtain $E_{elastic}(\theta _{1})$. By expanding 
$E_{elastic}(\theta _{1})$
about the minimum energy orbital state for $m=0$, $\theta _{1}(m=0)$, we
obtain 
\begin{equation}
E_{elastic}(\theta _{1})\approx \frac{1}{2}K_{\theta }\left( \theta
_{1}-\theta _{1}(m=0)\right) ^{2},
\end{equation}
where $K_{\theta }=12\lambda ^{2}(K^{*}-K_{2s})/(K_{2s}K^{*})$, 
and $\theta
_{1}(m=0)=\frac{1}{2}\cos ^{-1}[\sqrt{3}\lambda 
A/(K_{2s}(K^{*}-K_{2s}))]$.
By substituting $T_{c}(\theta _{1})\approx T_{c}^{0}[1+\alpha (\theta
_{1}-\theta _{1}(m=0))]$ and $E_{elastic}(\theta _{1})$ in Eq. (\ref
{eq:Landauf}), and minimizing $f(\theta _{1},m)$ about $\theta _{1}$, we
obtain 
\begin{equation}
\theta _{1}(m)-\theta _{1}(m=0)=\frac{1}{K_{\theta }}
\left[ 2T_{c}^{0}\alpha
m^{2}-T_{c}^{0}\alpha m^{4}\right] ,
\end{equation}
\begin{equation}
f(m)=2(T-T_{c}^{0})m^{2}+\frac{T_{c}^{0}}{K_{\theta }}(K_{\theta }-2\alpha
^{2}T_{c}^{0})m^{4}+O(m^{6}).
\end{equation}
Therefore, we will have the second order phase transition when $K_{\theta
}>2\alpha ^{2}T_{c}^{0}$, and the first order transition when $K_{\theta
}<2\alpha ^{2}T_{c}^{0}$. This result implies that when the lattice is 
soft
(small $K_{\theta }$) or the coupling between magnetic interaction and
lattice is large (large $\alpha $), the transition becomes first-order. 
The
estimate of the parameters for ${\rm LaMnO_{3}}$ in Sect. VI.C predicts 
the
second-order phase transition. Above equation for $\theta _{1}(m)$ also
gives the net change of $\theta _{1}$, $\theta _{1}(T=0)-\theta
_{1}(T>T_{c})=\alpha T_{c}^{0}/K_{\theta }$.

\section{Model of band structure and optical conductivity}

We use a tight-binding approximation to calculate band structure and 
optical
conductivity for strained films. This method is explained in detail in 
Ref. %
\onlinecite{Ahn.Millis.optic}, and summarized in this section. According 
to
band theory calculations\cite{Satpathy96.JAP,Satpathy96.PRL.Pickett96} the
conduction band is derived mainly from the Mn $e_{g}$ symmetric d-orbitals
and is well separated from other bands. Therefore, we only consider Mn $%
e_{g} $ levels. Kinetic energy and chemical potential terms are 
\begin{equation}
H_{{\rm KE}}+H_{\mu }=-\frac{1}{2}
\sum_{\vec{i},\vec{\delta},a,b,\alpha }t_{%
\vec{\delta}}^{ab}d_{\vec{i}a\alpha }^{\dagger }d_{\vec{i}+\vec{\delta}%
b\alpha }+H.c.-\mu 
\sum_{\vec{i},a,\alpha }d_{\vec{i}a\alpha }^{\dagger }d_{%
\vec{i}a\alpha }.
\end{equation}
Here $\vec{i}$ represents the coordinates of Mn sites, 
$\delta (=\pm x,y,z)$
labels the nearest neighbors of Mn sites, $a$ and $b$ represent the two
degenerate Mn $e_{g}$ orbitals on a site, $\alpha $ denotes the spin 
state,
and $t_{\vec{\delta}}^{ab}$ is the hopping amplitude between orbital $a$ 
on
site $\vec{i}$ and $b$ on site $\vec{i}+\vec{\delta}$. We choose $|\psi
_{1}>=|3z^{2}-r^{2}>$ and $|\psi _{2}>=|x^{2}-y^{2}>$ as the two linearly
independent $e_{g}$ orbitals on each site as before. The hopping 
matrix $t_{%
\vec{\delta}}^{ab}$ has a special form: For hopping along $z$ direction, 
it
connects only the two $|3z^{2}-r^{2}>$ states, thus $%
t_{z}^{ab}=t_{-z}^{ab}=t_{o}$ for a=b=1, and zero otherwise. The hopping
matrices in other bond directions are obtained by appropriate rotations. 
The
Jahn-Teller coupling for uniform $Q_{3}$ distortion and staggered $Q_{2}$
distortion with wave vector $\vec{K}_{lattice}=(\pi ,\pi ,0)$ or 
$(\pi ,\pi
,\pi )$ is 
\begin{equation}
H_{{\rm JT}}=-\sqrt{\frac{3}{2}}\lambda \sum_{\vec{i},\alpha }\left( 
\begin{array}{c}
d_{1,\vec{i},\alpha }^{\dagger } \\ 
d_{2,\vec{i},\alpha }^{\dagger }
\end{array}
\right) ^{T}\left( 
\begin{array}{cc}
-Q_{3u} & \exp (i\vec{K}_{lattice}\cdot \vec{i})Q_{2s} \\ 
\exp (i\vec{K}_{lattice}\cdot \vec{i})Q_{2s} & Q_{3u}
\end{array}
\right) \left( 
\begin{array}{c}
d_{1,\vec{i},\alpha } \\ 
d_{2,\vec{i},\alpha }
\end{array}
\right) .
\end{equation}
The Hund's coupling for antiferromagnetic core spin configuration with 
wave
vector $\vec{K}_{spin}$ ($(0,0,\pi )$ for A type AF, $(\pi ,\pi ,\pi )$ 
for
purely AF) is 
\begin{equation}
H_{{\rm Hund}}=J_{{\rm H}}S_{{\rm c}}\sum_{\vec{i},a}
\left[ \left( 1-\exp (i%
\vec{K_{spin}}\cdot \vec{i})\right) d_{\vec{i},a,\uparrow }^{\dagger }d_{%
\vec{i},a,\uparrow }+\left( 1+\exp (i\vec{K_{spin}}\cdot \vec{i})\right) 
d_{%
\vec{i},a,\downarrow }^{\dagger }d_{\vec{i},a,\downarrow }\right] .
\end{equation}
The total Hamiltonian is the sum of the terms considered so far. By
diagonalizing this in $k$ space, we can find the energy levels and
eigenstates.

Optical conductivity per volume due to the transitions between Mn $e_{g}$
levels, $\sigma $, can be calculated using the eigenstates and energy 
levels
found from the above Hamiltonian. Using the standard linear response 
theory,%
\cite{Millis90}$^{\text{,}}$\cite{Dagotto94} optical conductivity is given
by 
\begin{equation}
\sigma _{p}^{\lambda \nu }=-\frac{1}{i\omega N_{Mn}a_{0}^{3}}\sum_{n}
\frac{%
<0|J_{p\lambda }^{\dagger }|n><n|J_{p\nu }|0>}{\hbar \omega
-(E_{n}-E_{0})+i\epsilon },
\end{equation}
where $\epsilon $ is an infinitesimal and $J_{p}$ is given by $\hat{\vec{%
J_{p}}}=-\frac{iea_{0}}{2\hbar }\sum_{\vec{i},\vec{\delta},a,b,\alpha }
t_{%
\vec{\delta}}^{ab}
\vec{\delta}\left( d_{\vec{i}a\alpha }^{\dagger }d_{\vec{i}%
+\vec{\delta}b\alpha }-H.c.\right) $.\cite{Drude}

\section{Comparison of $(\pi ,\pi ,0)$ and $(\pi ,\pi ,\pi )$ type
distortions}

In the purely local model considered in Ref. \onlinecite{Millis96}, for
which only nearest-neighbor Mn-O and Mn-Mn springs are considered, we 
have $%
K_{s}$ = $K_{2s}$ and $E_{tot}^{MIN}(\pi \pi \pi )=E_{tot}^{MIN}
(\pi \pi 0)$
up to order $A^{2}$ according to Eqs. (\ref{eq:Eminpipi0}) and (\ref
{eq:Eminpipipi}). However, in real situation for which farther ion-ion
elastic energies exist, we have $K_{s}$ $\neq $ $K_{2s}$ and the two
distortions have different energies$.$ Equations (\ref{eq:Eminpipi0}) 
and (%
\ref{eq:Eminpipipi}) shows that if $K_{2s}<K_{s}$, then the leading order
term stabilizes the $(\pi ,\pi ,0)$ distortion over the $(\pi ,\pi ,\pi )$
distortion. If $K_{s}$ and $K_{2s}$ are very close to $K^{*}$, the $A^{2}$
term will be smaller for the larger of $K_{s}$ and $K_{2s}$, opposite 
trend
of the leading order term. For parameter values determined in 
Appendix B, we
find that $K_{2s}<K_{s}$ is required to make $(\pi \pi 0)$ distortion
favored. Optic phonon spectrum along $\vec{k}=\kappa (\pi ,\pi ,0)$ and $%
\kappa (\pi ,\pi ,\pi )$ ($0<\kappa <1$ ) would be useful to check this
condition. At $\kappa =0$, these two modes have the same energy. 
As $\kappa $
approaches to 1, if the $(\pi ,\pi ,\pi )$ mode has a higher energy 
than the 
$(\pi ,\pi ,0)$ mode, it would be an indication that $K_{s}>K_{2s}$. 
So far,
the phonon spectrum for ${\rm LaMnO_{3}}$ has not been calculated. Ghosez 
{\it et al.}\cite{Ghosez99} have calculated phonon spectra for similar
compounds, ${\rm BaTiO_{3}}$, ${\rm PbTiO_{3}}$, and ${\rm PbZrO_{3}}$.
These results show that $(\pi ,\pi ,0)$ mode ($M_{\text{2}}$ point 
in Fig. 1
of Ref. \onlinecite{Ghosez99}) has a higher energy than $(\pi ,\pi ,\pi )$
mode ($R_{\text{12'}}$ point in Fig.1. of Ref. \onlinecite{Ghosez99}) 
by 11
\%, 7\%, and 2 \% respectively, contrary to our expectation for ${\rm %
LaMnO_{3}}$. Since the energies of these modes depend sensitively on
transition metal elements, a phonon spectrum calculation for 
${\rm LaMnO_{3}}
$ is necessary.

We examine the possibility that the two distortions have different band
energies. For this purpose, we use a tight binding Hamiltonian 
introduced in
the previous section, and calculate the band structures and total band
energies for the two distortion patterns. On the first Brillouin zone
boundary (i.e., $|k_{z}|=\pi /2$ or $|k_{x}|+|k_{y}|=\pi $ planes) and on
the planes satisfying $|k_{x}|=|k_{y}|$, we find that the band structures
for $(\pi ,\pi ,0)$ and $(\pi ,\pi ,\pi )$ distortions are identical.
Between these planes, when a band is well separated from other bands, 
it has
a similar band structure for the $(\pi ,\pi ,0)$ and $(\pi ,\pi ,\pi )$
distortions. Since the filled bands are well separated from the 
empty bands
by the Jahn-Teller splitting, the results show that the change of the 
filled
bands are negligible. We find that the total band energy per Mn ion 
changes
less than 1 meV between the $(\pi ,\pi ,0)$ and $(\pi ,\pi ,\pi )$
distortions. We therefore conclude that the strain (lattice restoring 
force)
effects are crucial.

In our calculation for the strained films, we assume that the sign of the
energy difference between the two types of ordering is not changed by
applied strains, and consider the $(\pi ,\pi ,0)$ ordering only.

\section{Results}

\subsection{Lattice and orbital states in strained films}

In this section, we present our calculations for strained films. We use $%
e_{||}$ to denote the substrate induced additional strain, $%
e_{xx}-e_{xx}^{bulk}$. We examine the strain in the range of -2 \% $<$ $%
e_{||}$ $<$ 2 \%. 
Since $a_{0}\approx 4.03\stackrel{\circ }{\text{A}}$, the
difference of the $xy$ plane lattice parameter between 
bulk ${\rm LaMnO_{3}}$
and substrate is between $-0.08$ $\stackrel{\circ }{\text{A}}$ and 0.08 $%
\stackrel{\circ }{\text{A}}$. We represent the changes in $Q_{3u}$, 
$Q_{2s}$%
, and $e_{zz}$ by writing $\delta $ in front. For example, $\delta
Q_{3u}=Q_{3u}-Q_{3u}^{bulk}$. Our calculation shows that $\delta e_{zz}$
versus $e_{||}$ in this range is close to linear. We obtain 
$\delta e_{zz}$/$%
e_{||}$ $\approx $ -1.8 for the parameter set obtained from Ref. %
\onlinecite{Millis.strain}, and $\delta e_{zz}$/$e_{||}$ $\approx $ -1.4 
for
the parameter set obtained from Ref. \onlinecite{Japan99}. These ratios 
are
about 2 times larger than the ratio for 
${\rm La_{0.7}Ca_{0.3}MnO_{3}}$ film.%
\cite{Japan99} This is due to the softening of the $Q_{3u}$ mode 
in ${\rm %
LaMnO_{3}}$.

From $\delta e_{zz}$ versus $e_{||}$, we can also find $\delta Q_{1u}$ 
and $%
\delta Q_{3u}$ versus $e_{||}$. The results are shown in Fig. 2. To
understand these results, we find the leading terms of $\delta Q_{1u}$/($%
a_{0}e_{||}$), $\delta Q_{2s}$/($a_{0}e_{||}$), and $\delta Q_{3u}$/($%
a_{0}e_{||}$). They are 
\begin{eqnarray}
\frac{\delta Q_{1u}}{a_{0}e_{||}} &=&2\sqrt{3}\frac{K^{*}-K_{2s}}{%
K_{B}+2\left( K^{*}-K_{2s}\right) }+O(A^{2}), \\
\frac{\delta Q_{2s}}{a_{0}e_{||}} &=&-\frac{3}{\sqrt{2}}\frac{K_{B}\left(
2K^{*}-3K_{2s}\right) +4\left( K^{*}-K_{2s}\right) ^{2}}{\left(
K_{B}+2K^{*}-2K_{2s}\right) \left( K^{*}-K_{2s}\right) K_{2s}^{2}}\lambda
A+O(A^{3}), \\
\frac{\delta Q_{3u}}{a_{0}e_{||}} &=&-\sqrt{6}\frac{K_{B}}{K_{B}+2\left(
K^{*}-K_{2s}\right) }+O(A^{2}).
\end{eqnarray}

This shows that $Q_{2s}$ changes more slowly than $Q_{3u}$, because the
staggered distortion is coupled to the uniform strain only through the
anharmonic term. Since $K^{*}-K_{2s}$ is almost one order smaller than $%
K_{B} $, $Q_{1u}$ also changes slowly compared with $Q_{3u}$. 
Therefore, the
main effect of the uniaxial strain is the change in the uniform tetragonal
distortion without much change in staggered distortion or volume. Figure 2
shows that $\delta Q_{2s}$/($a_{0}e_{||}$) can be either positive or
negative depending on the parameter values, whereas $\delta Q_{1u}$/($%
a_{0}e_{||}$) $>$ 0 and $\delta Q_{3u}$/($a_{0}e_{||}$) $<$ 0 always.

We obtain $\theta _{1}\approx $ 54.8 $\stackrel{\circ }{\text{A}}$ for the
bulk state from Eq. (\ref{eq:th}). $\theta _{1}$ versus $e_{||}$ is 
shown in
Fig. 3. The change in $\theta _{1}$ is about $\pm 5^{o}-15^{o}$. 
For tensile
strains, $\theta _{1}$ and $\theta _{2}$ approach towards $90^{o}$, which
corresponds to $|x^{2}-y^{2}>$. For compressive strains, $\theta _{1}$ 
and $%
\theta _{2}$ approach to $0^{o}$ and $180^{o}$, which corresponds to $%
|3z^{2}-r^{2}>$. This can be understood from the fact that in 
$\theta _{1}$,$%
\theta _{2}$=$|x^{2}-y^{2}>$ state, the x-y plane Mn-O-Mn distance 
tends to
be farthest, and in $\theta _{1}$, $\theta _{2}$ = $|3z^{2}-r^{2}>$,
shortest due to the electron distribution in the $xy$ plane.

The change in the Jahn-Teller splitting, $\delta \Delta E_{JT}$, is 
shown in
Fig. 4, which is about $\pm $ 0.02-0.2 eV for $\pm $ 2 \% strain.

\subsection{Magnetic property in strained films}

According to Ref. \onlinecite{Moussa96}, $J_{xy}$ = -1.66 meV, 
$J_{z}$=1.16
meV for bulk ${\rm LaMnO_{3}}$. Since $\theta _{1}=54.8^{o}$ for bulk, we
obtain $J_{F}=7.52$ meV and $J_{AF}=4.50$ meV, which corresponds to $%
J_{xy}^{0}=-0.22$ meV, $J_{z}^{0}=2.62$ meV, and $J^{*}=1.88$ meV. Using
these parameter values, we find $J_{xy}$ and $J_{z}$ versus $\theta _{1}$,
which are plotted in Fig. 5 (a). As the two orbital states approach 
to $\pi
/4$ and $3\pi /4$, i.e., $|z^{2}-x^{2}>$ and $|z^{2}-y^{2}>$, 
both $J_{xy}$
and $J_{z}$ become more ferromagnetic, and as they approach to 0 
and $\pi $,
i.e., $|3z^{2}-r^{2}>$ and $|3z^{2}-r^{2}>$, or to $\pi /2$ and $\pi /2$,
i.e., $|x^{2}-y^{2}>$ and $|x^{2}-y^{2}>$, $J_{xy}$ and $J_{z}$ 
become less
ferromagnetic. This is due to the orbital-state-dependent hopping between
the filled and empty orbitals, which mediates ferromagnetic interaction.
When $20^{o}<\theta _{1}<70^{0}$, the magnetic ground state remains A-type
antiferromagnetic. Outside this range, both $J_{xy}$ and $J_{z}$ become
positive, and purely antiferromagnetic state is the ground state. In fact,
Fig. 3 shows that 2 \% tensile strain can change $\theta _{1}$ close to $%
70^{o}$, and turn the material into a purely antiferromagnetic state.

The mean field estimates of $T_{c}$ are shown in Fig. 5 (b). $T_{c}$ for
bulk state is about 210 K, somewhat larger than the measured $T_{c}$ = 
140 K.%
\cite{Moussa96} It shows that $\pm $ 2 \% strain changes $T_{c}$ by 
about $%
\pm $ 50 K. Relatively large change in $T_{c}$ and the possible 
change into
purely AF state are due to the strong dependence of magnetism 
on the $e_{g}$
orbital state and the strong Jahn-Teller coupling between the $e_{g}$
orbital state and the lattice distortion.

\subsection{Magnetostriction effects in bulk state}

In this subsection we present the magnetostriction effects calculated 
by the
model in Section III. B. We use the two sets of parameter values in 
Table I
and $J^{*}=$ 1.88 meV. The results obtained by numerical minimization 
of the
Landau free energy [Eq. (\ref{eq:Landauf})] are shown in Table II. The
change in the JT strain, $\delta \epsilon ^{*}$, is about 0.003 $\sim $
0.01. $\delta \epsilon ^{*}/[\epsilon ^{*}(T>T_{c})]$ is about -0.08 
for the
parameter set from Ref. \onlinecite{Millis.strain}, and -0.31 for 
that from
Ref. \onlinecite{Japan99}. These results show that when the effective JT
modulus of $Q_{3u}$ mode, $K^{*}-K_{2s}$, is smaller, the magnetostriction
effect is larger.

We obtain $\alpha \approx $2, and $2\alpha ^{2}T_{c}^{0}\approx $ 
0.1 eV for
both parameter sets from Refs. \onlinecite{Millis.strain} and %
\onlinecite{Japan99}. We obtain $K_{\theta }$ =1.2 eV for Ref. %
\onlinecite{Millis.strain}, and $K_{\theta }$ =0.3 eV for Ref. %
\onlinecite{Japan99}. Therefore, the transition will be of the 
second order.
However, if we have a softer $K_{\theta }$ (a third or a tenth), or a
stronger magnetostriction coupling $\alpha $ (twice or three times), 
then we
will have a first-order phase transition. Our numerical minimization 
of the
free energy confirms these results.

Recently, the orbital ordering in ${\rm LaMnO_{3}}$ has been directly
observed using a resonant X-ray scattering technique.\cite{Murakami98} In
this result, the orbital ordering versus temperature curve has a change of
the slope at $T=T_{N}$. The sign of the change indicates that the orbital
states change away from $3x^{2}-r^{2}/3y^{2}-r^{2}$ 
($\theta _{1}$=$60^{o}$)
as $T\rightarrow 0$ below $T_{N}$, which is consistent with our 
calculation.
Recent neutron diffraction study measured $\sin \theta _{1}$ 
versus $T$.\cite
{Carvajal98} Our results predict that $\sin \theta _{1}$ 
(which corresponds
to $c_{2}$ in Fig. 3 in Ref. \onlinecite{Carvajal98}) changes by 0.01 for
parameters from Ref. \onlinecite{Millis.strain} and 0.04 for parameters 
from
Ref. \onlinecite{Japan99} between T=0 and T $>T_{N}$. However, Ref. %
\onlinecite{Carvajal98} shows negligible change in $\sin \theta _{1}$
between
T=0 and T $>T_{N}$, which indicates a smaller magnetostriction coupling $%
\alpha $, or a larger elastic modulus $K_{\theta } $ than the values
obtained above.

\subsection{Band structures in strained films}

First, we summarize the band structure in bulk state A-type
antiferromagnetic ${\rm LaMnO_{3},}$ which is explained in 
detail in Ref. %
\onlinecite{Ahn.Millis.optic}. Crudely speaking, the bands fall into 4
pairs, which may be understood by setting $t_{o}=0$ [as occurs 
at $\vec{k}%
=(\pi /2,\pi /2,\pi /2)$]; in this case we have four separate energy 
levels
on each site, which are $E_{1,2}=-\sqrt{\frac{3}{2}}\lambda \sqrt{%
Q_{2s}^{2}+Q_{3u}^{2}}$, $E_{3,4}=\sqrt{\frac{3}{2}}\lambda \sqrt{%
Q_{2s}^{2}+Q_{3u}^{2}}$, 
$E_{5,6}=2J_{{\rm H}}S_{{\rm c}}-\sqrt{\frac{3}{2}}%
\lambda \sqrt{Q_{2s}^{2}+Q_{3u}^{2}}$, and 
$E_{7,8}=2J_{{\rm H}}S_{{\rm c}}+%
\sqrt{\frac{3}{2}}\lambda \sqrt{Q_{2s}^{2}+Q_{3u}^{2}}$. To find the three
parameter values of our model Hamiltonian, i.e., $t_{0}$, $\lambda $, 
and $%
J_{{\rm H}}S_{{\rm c}}$, we fit our band structure calculation to the LDA
(local density-functional approximation) band calculation for the JT
distorted ${\rm LaMnO_{3}}$ in Ref.\onlinecite{Satpathy96.JAP} at high
symmetry points in reciprocal space. The standard deviation is $\approx $
0.2 eV. The determined parameter values are $t_{o}$=0.622 eV, $\lambda $%
=1.38 eV/$\stackrel{\circ }{\text{A}}$, and $2J_{{\rm H}}S_{{\rm c}}$=2.47
eV. The fitted band structure is shown in Fig. 1 in Ref. %
\onlinecite{Ahn.Millis.optic}.

When the strain does not change A-type antiferromagnetic core spin
configuration, the main effect of the strain is the change in the band
width. The results are shown in Fig. 6. Solid lines are for the 
compressive
strain, and dashed lines are for the tensile strain. Bulk band 
structure can
be approximately obtained by taking the average of the two band 
structures.
At $(\pi /2,\pi /2,\pi /2)$, where the effective hopping vanishes, the
energy level change is $\delta \Delta E_{JT}$ obtained before. From this
point, dispersions along $(\pi /2,\pi /2,0)$ and along $(\pi ,0,\pi /2)$
represent the hoppings in $z$ direction and in $xy$ plane, respectively.
Between $(\pi /2,\pi /2,\pi /2)$ and $(\pi /2,\pi /2,0)$, the 
widths of the
lower JT bands, $E_{1,2}$ and $E_{5,6}$, are increased for compressive
strains, whereas the widths of the upper JT bands, $E_{3,4}$ and 
$E_{7,8}$,
are decreased. This is related to the changes in $\theta _{1}$ and $\theta
_{2}$ due to the strains: As $\theta _{1}$ and $\theta _{2}$ approach to 0
and $\pi $, the lower JT level approaches to $3z^{2}-r^{2}$ state 
which has
a large hopping along $z$ direction, whereas the upper JT level approaches
to $x^{2}-y^{2}$ which has no hopping along $z$ direction. 
But between $(\pi
/2,\pi /2,\pi /2)$ and $(\pi ,0,\pi /2)$, the dispersion does not change
much, indicating that the average hopping is not changed much in $xy$ 
plane
due to the alternating orbital pattern in $xy$ plane.

As pointed out in the previous section, 2 \% tensile strain may induce
purely antiferromagnetic ground state. Due to the strong Hund's coupling,
this results in substantial changes in band structure and optical
conductivity as shown below and in the next section, respectively. Band
structure for the purely AF state can be obtained by using $\vec{K}%
_{spin}=(\pi ,\pi ,\pi )$ in the model described in Sect. IV. The results
are shown in Fig. 7 for the same lattice distortions and parameters 
used for
the band structure shown as dotted lines in Fig. 6. Between $(\pi /2,\pi
/2,\pi /2)$ and $(\pi /2,\pi /2,0)$, the two band structures are identical
since it involves only $z$ direction hopping. Between $(\pi /2,\pi /2,\pi
/2) $ and $(\pi ,0,\pi /2)$, where only $xy$ directional hopping is
involved, for A-type AF state the different JT levels repel each other,
while for purely AF state the different Hund's levels repel each other. 
This
represents different mixings for different spin configurations: the mixing
for A-type AF state is mainly between the different JT states, while the
mixing for purely AF state is mainly between the different Hund's states.
the band structure between $(\pi ,0,0)-(0,0,0)-(\pi /2,\pi /2,\pi /2)$ 
shows
the same trend, which results in a small indirect band gap for 
the purely AF
state. However, due to the on-site Coulomb repulsion of about 1-2 eV
neglected in the above calculation\cite{Ahn.Millis.optic} it is 
unlikely to
have the insulator-to-semimetal transition in this material.

\subsection{Optical conductivities in strained films}

From the band structure, we have calculated optical conductivities for
strained films in A-type AF ground state. Results are shown in Fig. 8. $%
\sigma _{xx}$ or $\sigma _{yy}$ (solid lines) shows a relatively small
changes by strains. The spectral weight of the Hund's peak 
in $\sigma _{zz}$
(dotted lines) at around 2.5 eV is increased (decreased) about 20 \% as we
apply 2 \% compressive (tensile) strain. This seems consistent with the
change of the average hopping along $z$ direction by strains mentioned in
the previous section.

Optical conductivities for the purely AF state are calculated in the same
way. Figure 9 shows the results for $\sigma _{xx}$ (solid line) 
and $\sigma
_{zz}$ (dotted line). First, the sharp Hund's peak in $\sigma _{zz}$ is
disappeared. This can be understood from the band structure, particularly
between $(\pi /2,\pi /2,\pi /2)$ and $(\pi ,0,\pi /2)$. The sharp Hund's
peak in $\sigma _{zz}$ for A-type AF state originates from the two 
parallel
bands split by the Hund's coupling, 2$J_{\text{H}}S_{\text{c}}$. However,
this structure disappears when we have purely AF core spin as seen in Fig.
7. Comparison of Fig. 9 (a) (2 \% strain) and (b) (bulk) shows that as
orbital state is changed toward $x^{2}-y^{2}$ by the tensile strain, which
has zero hopping along z-direction, the spectral weight of the Hund's peak
in $\sigma _{zz}$ decreases, as observed for the A-type AF state 
in Fig. 8. $%
\sigma _{xx}$ shows prominent peaks at the Hund's 
splitting (2.5-3.5 eV) and
at the Hund-plus-JT splitting (4-5 eV) due to purely AF spin state in
contrast to the A-type AF state (see Fig. 8). The JT peaks at 
around 1 eV in
both $\sigma _{xx}$ and $\sigma _{zz}$ are due to the strong hybridization
between major and minor spin states. If we increase $J_{H}S_{c}$ and 
reduce
this hybridization, the JT peaks decrease, as can be seen by 
comparing Figs.
9(a) and 9(c). Above results show that changes in $xy$ plane spin
configuration make differences in $\sigma _{zz}$ due to the changes in
hybridization.

For the calculations so far, we have assumed zero on-site 
Coulomb repulsion $%
U$. However, as explained in Ref. \onlinecite{Ahn.Millis.optic}, in this
material there exists $U\approx 1.6$ eV. Therefore actual peak positions
will be higher by $\sim $ 1.6 eV and the spectral weight of each peak will
be reduced inversely proportional to the peak energy.

\section{Conclusion}

In summary, 
we developed a model of elastic energy for ${\rm LaMnO_{3}}$ and
solved for uniaxial strains in thin films. We found that $\pm 2$ \% strain
can change the uniform tetragonal strain and $e_{g}$ orbital states 
without
much change in the staggered distortion or volume. We found that 2 \%
tensile strain can change the magnetic ground state into purely
antiferromagnetic state, inducing dramatic changes in band structure and
optical conductivity. Magnetostriction effect at $T_{\text{N}}$ in bulk
state is found to be large. We examined the possibility that the lattice
energy will favor ($\pi \pi 0$) ordering over ($\pi \pi \pi $) ordering. 
We
also noted ``$Q_{3u}$ mode softening'' in ${\rm LaMnO_{3}}$. The results
presented in this paper for ${\rm LaMnO_{3}}$ show the strong coupling
between lattice, Mn $e_{g}$ orbital state, and exchange interaction, which
lies at the root of the novel properties of doped manganese perovskites.

This work is supported by NSF-DMR-9705182 and the University of Maryland
MRSEC.

\appendix

\section{General expression of elastic energy for staggered distortion}

In this Appendix, we show how we can get the general expression of the
elastic energy due to distortions with wave vector $\vec{k}$ in perovskite
structure. 
We again consider three dimensional displacement of Mn ion, $\vec{%
\delta}$, and the displacements of oxygen ions along Mn-Mn axis, 
$u^{x}$, $%
u^{y}$, and $u^{z}$ as defined in the text and shown in Fig. 1. 
We consider
the displacements with wave vector $\vec{k}$, 
$\vec{\delta}_{\vec{i}}=\vec{%
\delta}_{\vec{k}}e^{i\vec{k}\cdot \vec{i}}$ and 
$u_{\vec{i}}^{x,y,z}=u_{\vec{%
k}}^{x,y,z}e^{i\vec{k}\cdot \vec{i}}$. When we define $d_{1,2,3}(\vec{k}%
)=\delta _{\vec{k}}^{x,y,z}$ and $d_{4,5,6}(\vec{k})=u_{\vec{k}}^{x,y,z}$,
the energy due to these strains, $E_{s}(\vec{k})$, is given by 
\begin{equation}
E_{s}(\vec{k})=\sum_{i,j}d_{i}(\vec{k})D_{ij}(\vec{k})d_{j}(\vec{k}),
\end{equation}
where $D_{ij}(\vec{k})=D_{ji}(\vec{k})$.

In special cases, symmetry arguments make certain terms vanish or equal.
First, when $k_{x}=\pi $, mirror symmetry operation about $x=0$ plane
changes $k_{x}$ to $-k_{x}$. Since $k_{x}=\pi $ and 
$k_{x}^{\prime }=-\pi $
are equivalent, the symmetry operation changes $\vec{k}$ back to 
$\vec{k}$.
Since this operation changes $\delta _{\vec{k}}^{x}$ to 
$-\delta _{\vec{k}%
}^{x}$, odd order terms of $\delta _{\vec{k}}^{x}$ should vanish, and
therefore, 
$D_{12}(\vec{k})=D_{13}(\vec{k})=D_{14}(\vec{k})=D_{15}(\vec{k}%
)=D_{16}(\vec{k})=0$. Second, when $k_{x}=0$, mirror symmetry operation
about $x=0$ plane changes $\vec{k}$ back to $\vec{k}$, $\delta _{\vec{k}%
}^{x} $ to $-\delta _{\vec{k}}^{x}$, and $u_{\vec{k}}^{x}$ to 
$-u_{\vec{k}%
}^{x}$. 
Therefore, odd order terms of $\delta _{\vec{k}}^{x}$ and $u_{\vec{k}%
}^{x}$ vanish. Therefore, $D_{12}(\vec{k})=D_{13}(\vec{k})=D_{15}(\vec{k}%
)=D_{16}(\vec{k})=0$ and $D_{42}(\vec{k})=D_{43}(\vec{k})=D_{45}(\vec{k}%
)=D_{46}(\vec{k})=0$. Third, when $k_{x}=\pm k_{y}$, mirror operation
interchanging $x$ axis and $\pm y$ axis changes $\vec{k}$ 
back to $\vec{k}$,
and 
$\delta _{x}$ to $\pm \delta _{y}$, $u_{x}$ to $\pm u_{y}$. Therefore, $%
D_{11}(\vec{k})=
D_{22}(\vec{k})$, $D_{44}(\vec{k})=D_{55}(\vec{k})$, $D_{14}(%
\vec{k})=D_{25}(\vec{k})$, $D_{15}(\vec{k})=D_{24}(\vec{k})$, 
$D_{1j}(\vec{k}%
)=\pm D_{2j}(\vec{k})$, $D_{4j}(\vec{k})=\pm D_{5j}(\vec{k})$, 
where $j=3,6$

If we apply these rules to $\vec{k}=(\pi ,\pi ,0)$, and $(\pi ,\pi ,\pi )$
distortions, we obtain the following expressions. 
\begin{eqnarray}
\frac{E_{s}[\vec{k}=(\pi ,\pi ,0)]}{N_{Mn}} &=&D_{11}(\vec{k})({\delta _{%
\vec{k}}^{x}}^{2}+{\delta _{\vec{k}}^{y}}^{2})+
D_{33}(\vec{k}){\delta _{\vec{%
k}}^{z}}^{2}+D_{44}(\vec{k})({u_{\vec{k}}^{x}}^{2}+{u_{\vec{k}}^{y}}^{2}) 
\nonumber \\
&&+D_{66}(\vec{k}){u_{\vec{k}}^{z}}^{2}+
2D_{45}(\vec{k})u_{\vec{k}}^{x}u_{%
\vec{k}}^{y}+2D_{36}(\vec{k})\delta _{\vec{k}}^{z}u_{\vec{k}}^{z}.
\label{Es}
\end{eqnarray}
\begin{eqnarray}
\frac{E_{s}[\vec{k}=(\pi ,\pi ,\pi )]}{N_{Mn}} &=&D_{11}
(\vec{k})({\delta _{%
\vec{k}}^{x}}^{2}+{\delta _{\vec{k}}^{y}}^{2}+{\delta _{\vec{k}}^{z}}%
^{2})+D_{44}(\vec{k})({u_{\vec{k}}^{x}}^{2}+{u_{\vec{k}}^{y}}^{2}+
{u_{\vec{k}%
}^{z}}^{2})  \nonumber \\
&&+2D_{45}(\vec{k})(u_{\vec{k}}^{x}u_{\vec{k}}^{y}+
u_{\vec{k}}^{y}u_{\vec{k}%
}^{z}+u_{\vec{k}}^{z}u_{\vec{k}}^{x}).
\end{eqnarray}

$K_{2s}$, $K_{3s}$, and $K_{s}$ in the text are defined as $%
K_{2s}=[D_{44}(\pi \pi 0)-D_{45}(\pi \pi 0)]/2$, $K_{3s}=3[D_{44}(\pi \pi
0)+D_{45}(\pi \pi 0)]/2$, and $K_{s}=[D_{44}(\pi \pi \pi )-D_{45}(\pi \pi
\pi )]/2$. Therefore, the condition for $K_{3s}>K_{2s}$ is $D_{44}(\pi \pi
0)>-2D_{45}(\pi \pi 0)$. We expect this condition is well-satisfied 
for the
following reasons: First, the stability of lattice implies $D_{44}(\pi \pi
0)>0$. Second, the Coulomb repulsion between oxygen will favor similar O-O
distances, which implies $D_{45}(\pi \pi 0)>0$.

\section{Parameters for the model of elastic energy}

Optical absorption experiment for ${\rm LaMnO_{3}}$ shows Mn-O bond
stretching mode peak at 70.3 meV.\cite{Jung98} From this we can find the
effective Mn-O spring constant 
$K_{1}$=7.36 eV/$\stackrel{\circ }{\text{A}}%
^{2}$ (Ref. \onlinecite{Ahn.Millis.optic}). If we assume 
that $K_{1}$ is the
main contribution 
to $K_{2s}$, we obtain $K_{2s}\approx K_{1}$/2 = 3.68 eV/$%
\stackrel{\circ }{\text{A}}^{2}$. 
In Ref. \onlinecite{Ahn.Millis.optic}, we
used this value of $K_{1}$ to estimate the Jahn-Teller coupling 
constant $%
\lambda $, the value of which is consistent with LDA band 
calculation.\cite
{Ahn.Millis.optic}

For ${\rm La_{0.83}Sr_{0.17}MnO_{3}}$, $c_{11}$ and $c_{12}$ have been
measured in an ultrasound experiment.\cite{Darling98} We believe the bulk
modulus 
$c_{B}=(c_{11}+2c_{12})/3$ =143 GPa($K_{B}$=10.8 eV/$\stackrel{\circ 
}{\text{A}}^{2}$) of ${\rm La_{0.83}Sr_{0.17}MnO_{3}}$ at 200 K
(orthorhombic structure) can be used as an approximate value for the bulk
modulus of ${\rm LaMnO_{3}}$. The value of $c_{B}$ at 310 K (rhombohedral
phase) is about 176 GPa, which gives a rough estimate of uncertainty in $%
c_{B}$ of about 20 \%.

Since ${\rm La_{0.83}Sr_{0.17}MnO_{3}}$ is in the doping region where a
structural change happens,\cite{Urushibara95} and $c^{*}$ is sensitive to
the structural transition unlike bulk modulus, we do not believe the
measured $c^{*}$ of this material is a good estimate 
of $c^{*}$ for ${\rm %
LaMnO_{3}}$. Indeed, $c^{*}$ for ${\rm La_{0.83}Sr_{0.17}MnO_{3}}$ is much
smaller than those for other typical perovskite oxides.\cite{King-Smith94}
For example, the LDA calculation in Ref.\onlinecite{King-Smith94} 
predicted $%
c_{B}=$ 199 GPa, $c^{*}=$ 142 GPa, and $c^{*}/c_{B}=$ 0.71 for ${\rm %
SrTiO_{3}}$, which is close to the measurements for ${\rm SrTiO_{3}}$, $%
c_{B} $ =179 GPa, $c^{*}$=115 GPa, and $c^{*}/c_{B}$ = 0.64. The same LDA
calculation results showed that other perovskite oxides, such 
as BaTiO$_{3}$%
, CaTiO$_{3}$, KNbO$_{3}$, NaNbO$_{3}$, PbTiO$_{3}$, PbZrO$_{3}$, 
BaZrO$_{3}$
However, for ${\rm %
La_{0.83}Sr_{0.17}MnO_{3}}$, $c_{B}$=143 GPa, $c^{*}$=48 GPa , and $%
c^{*}/c_{B}$=0.34 at 200 K and $c_{B}$=176 GPa, $c^{*}$=35 GPa , and $%
c^{*}/c_{B}$=0.20 at 300 K. Therefore $c^{*}/c_{B}$ ratio is 
less than half
of the values for typical perovskites.

Therefore, we instead use the $c_{12}/c_{11}$ ratio measured in ${\rm %
La_{0.7}Ca_{0.3}MnO_{3}}$ thin film (Ref. \onlinecite{Millis.strain}) 
and $%
{\rm La_{0.6}Sr_{0.4}MnO_{3}}$ thin film (Ref. \onlinecite{Japan99}), 
since
their doping ranges are relatively far from the structural phase 
transition
doping ratios. The results are $c_{12}/c_{11}$=0.312 for Ref. %
\onlinecite{Millis.strain} and $c_{12}/c_{11}$=0.374 for Ref. %
\onlinecite{Japan99}. Using 
\begin{equation}
\frac{c^{*}}{c_{B}}=\frac{3}{2}\frac{1-c_{12}/c_{11}}{1+2c_{12}/c_{11}},
\end{equation}
we obtain $c^{*}$ and $K^{*}$ shown in Table I.

We obtain $Q_{2s}^{eq}$ and $Q_{3u}^{eq}$ for bulk ${\rm LaMnO_{3}}$ from
crystallography data: 
$Q_{2s}^{eq}$=0.398 $\stackrel{\circ }{\text{A}}$, $%
Q_{3u}^{eq}$=-0.142 $\stackrel{\circ }{\text{A}}$ (Ref. %
\onlinecite{Ellemans71}). Therefore, the three unknown quantities of the
model, $Q_{1u}^{eq}$, $\lambda $, and $A$, are determined from the
equilibrium condition: 
\begin{equation}
\left. \frac{\partial E}{\partial Q_{1u}}\right| _{eq}=0,\left. \frac{%
\partial E}{\partial Q_{2s}}\right| _{eq}=0,\left. \frac{\partial E}{%
\partial Q_{3u}}\right| _{eq}=0.
\end{equation}
Obtained parameter values are shown in Table I. The values 
of $A$ are small
enough to justify our approximation of anharmonic terms. For example, the
largest dropped term, $AQ_{3u}^{3}/(3\sqrt{2})$, is 0.5-2 \% of $%
K^{*}Q_{3u}^{2}/2$ for parameters in Table I. We obtain 
$Q_{1u}^{\text{eq}}=$
0.024 $\stackrel{\circ }{\text{A}}$ and 0.005 
$\stackrel{\circ }{\text{A}}$
for the parameters from Refs. \onlinecite{Millis.strain} and %
\onlinecite{Japan99}, respectively, which shows that the average 
bond length
does not change much. Therefore, it is reasonable to 
approximate $a_{0}$ by
the average Mn-Mn distance observed in bulk ${\rm LaMnO_{3}}$, 4.03 $%
\stackrel{\circ }{\text{A}}$. The $\lambda $ values are close to the
independent estimate $\lambda $=1.38 eV/$\stackrel{\circ }{\text{A}}$
obtained from band structure fitting. Negative sign of $A$ is consistent
with thermal expansion. It is noteworthy that the size of $A$ is largely
different for the two parameter sets, which is the consequence of the $%
Q_{3u} $ mode softening: a small difference in $K^{*}$ gives a quite large
difference in $K^{*}-K_{2s}$ (0.89 eV/$\stackrel{\circ }{\text{A}}^{2}$and
0.19 eV/$\stackrel{\circ }{\text{A}}^{2}$for the parameters from Refs. %
\onlinecite{Millis.strain} and \onlinecite{Japan99}, respectively), which
results in a large difference in the estimate of $A$.

\newpage

\begin{figure}[tbp]
\caption{ Degrees of freedom considered in our model, 
$\vec{\delta}_{\vec{i}%
} $ and $u_{\vec{i}}^{x,y,z}$. }
\label{fig1}
\end{figure}

\begin{figure}[tbp]
\caption{$\delta Q_{1u}$, $\delta Q_{2s}$, $\delta Q_{3u}$ versus $e_{||}$
for (a) $K_B$=10.8 eV/$\stackrel{\circ }{\text{A}}^2$, $K^*$=3.87 eV/$%
\stackrel{\circ }{\text{A}}^2$, 
$\lambda$=1.25 eV/$\stackrel{\circ }{\text{A}%
}$, $K_{2s}$=3.68 eV/$\stackrel{\circ }{\text{A}}^2$, $A$=-0.329 eV/$%
\stackrel{\circ }{\text{A}}^3$, which are the parameter values 
from Ref. 2,
and (b) $K_B$=10.8 eV/$\stackrel{\circ }{\text{A}}^2$, $K^*$=4.57 eV/$%
\stackrel{\circ }{\text{A}}^2$, 
$\lambda$=1.13 eV/$\stackrel{\circ }{\text{A}%
}$, $K_{2s}$=3.68 eV/$\stackrel{\circ }{\text{A}}^2$, $A$=-1.65 eV/$%
\stackrel{\circ }{\text{A}}^3$, which are parameter values from Ref. 20,
shown in Table I.}
\label{fig2}
\end{figure}

\begin{figure}[tbp]
\caption{$\theta_1$ versus $e_{||}$: The solid line is for the parameter
values from Ref. 2, and the dashed line is for the parameter values from
Ref. 20, shown in Table I. }
\label{fig3}
\end{figure}

\begin{figure}[tbp]
\caption{$\delta E_{JT}$ versus $e_{||}$. Solid line is for the parameter
values from Ref. 2, and dashed line is for the parameter values from Ref.
20, shown in Table I.}
\label{fig4}
\end{figure}

\begin{figure}[tbp]
\caption{ (a) Magnetic coupling constant $J_{xy}$ (solid line) and $J_z$
(dotted line) versus orbital state $\theta_1$. (b) Mean field 
estimation of $%
T_{c}$ versus orbital state $\theta_1$. the ground state is A type
antiferromagnetic between $20^o$ and $70^o$, and purely antiferromagnetic
outside this range. }
\label{fig5}
\end{figure}

\begin{figure}[tbp]
\caption{ Band structures for 2 \% compressive strain (solid lines) 
and 2 \%
tensile strain (dotted lines) for the parameter values from Ref. 2, 
shown in
Table I. Tight binding model parameter values are $t_0$=0.622 eV, 
2$J_HS_c$%
=2.47 eV, and $\lambda $=1.13 eV. A-type antiferromagnetic ground state is
assumed. }
\label{fig6}
\end{figure}

\begin{figure}[tbp]
\caption{ Band structure for purely antiferromagnetic core spin 
state for 2
\% strain for the parameter values from Ref. 2 shown in Table I, to be
compared with the dotted lines in Fig. 6. The tight binding parameter 
values
are $t_0$=0.622 eV, 2$J_HS_c$=2.47 eV, and $\lambda$=1.13 eV. }
\label{fig7}
\end{figure}

\begin{figure}[tbp]
\caption{ Optical conductivities, $\sigma_{xx}$ (solid lines) and $%
\sigma_{zz} $ (dotted lines), for 2 \% compressive strain (top panels), 
bulk
(middle panels), and 2 \% tensile strain (bottom panels) for the parameter
values from Ref. 2 (left panels) and for the parameter values from Ref. 20
(right panels) shown in Table I. A-type antiferromagnetic core 
spin state is
assumed. }
\label{fig8}
\end{figure}

\begin{figure}[tbp]
\caption{ Optical conductivities, $\sigma_{xx}$ (solid lines) and $%
\sigma_{zz}$ (dotted lines), for purely antiferromagnetic core spin
configuration with (a) 2 \% tensile strain, (b) bulk, and (c) a 
larger Hund
coupling and 2 \% tensile strain. Lattice distortions in Fig. 1 (a), 
$t_0$%
=0.622 eV, 2$J_HS_c$=2.47 eV, and $\lambda$=1.13 eV are used for the
calculation. }
\label{fig9}
\end{figure}

\newpage

\begin{table}[tbp]
\caption{ Obtained values of parameters: $c_B$ and $K_B$ are from Ref. 24,
and $K_{2s}$ is from the optic phonon mode frequency. We use two
experimental results in Refs. 2 and 20 to obtain the two sets of 
parameters
shown here. Details about how these numbers are obtained are shown in
Appendix B. }
\label{table1}
\begin{tabular}{c||cc|cccccc}
Source & $c_B$(GPa) 
& $c^*$(GPa) & $K_B$(eV/$\stackrel{\circ }{\text{A}}^2$)
& $K^*$(eV/$\stackrel{\circ }{\text{A}}^2$) 
& $K_{2s}$(eV/$\stackrel{\circ }{%
\text{A}}^2$) & $\lambda$(eV/$\stackrel{\circ }{\text{A}}$) & $A$(eV/$%
\stackrel{\circ }{\text{A}}^3$) 
& $K^*-K_{2s}$(eV/$\stackrel{\circ }{\text{A}%
}^2$) \\ \hline
Ref. \onlinecite{Millis.strain} & 143 & 90.8 & 10.8 
& 4.57 & 3.68 & 1.13 & 
-1.65 & 0.89 \\ 
Ref. \onlinecite{Japan99} & 143 & 76.8 & 10.8 & 3.87 
& 3.68 & 1.13 & -0.329
& 0.19
\end{tabular}
\end{table}

\begin{table}[tbp]
\caption{ Magnetostriction effect (numerical calculation) }
\label{table2}
\begin{tabular}{c|c||ccccc}
T & parameters & $Q_{1u}$($\stackrel{\circ }{\text{A}}$) & $Q_{2s}$($%
\stackrel{\circ }{\text{A}}$) 
& $Q_{3u}$($\stackrel{\circ }{\text{A}}$) & $%
\theta_1$ & $\epsilon^*$ \\ \hline
T=0 & from Ref. \onlinecite{Millis.strain} & 0.0246 & 0.401 & -0.131 & $%
53.6^o$ & -0.0328 \\ 
& from Ref. \onlinecite{Japan99} & 0.0051 & 0.411 & -0.099 & $51.5^o$ & 
-0.0248 \\ \hline
$T>T_c$ & from Ref. \onlinecite{Millis.strain} 
& 0.0244 & 0.399 & -0.142 & $%
54.8^o$ & -0.0355 \\ 
& from Ref. \onlinecite{Japan99} & 0.0049 & 0.399 & -0.143 & $54.8^o$ & 
-0.0358 \\ \hline
Differences between & from Ref. \onlinecite{Millis.strain} & 2.5 $\times
10^{-4}$ & 0.0024 & 0.011 & $-1.2^o$ & 0.0028 \\ 
$T=0$ and $T>T_c$ & from Ref. \onlinecite{Japan99} 
& 2.6 $\times 10^{-4}$ & 
0.012 & 0.044 & $-3.4^o$ & 0.011
\end{tabular}
\end{table}


\begin{references}
\bibitem{Es}  This expression can be written in more symmetric way. 
For the
staggered distortion with wave vector $\stackrel{\rightarrow }{\kappa }$,

\[
E_{s}(\stackrel{\rightarrow }{\kappa })=\frac{1}{2}\stackrel{\sim }{Q}%
^{T}(K_{av}I+\stackrel{\rightarrow }{\Delta K_{\stackrel{\rightarrow }{%
\kappa }}}\cdot \stackrel{\rightarrow }{\tau })\stackrel{\sim }{Q},
\]

where $\stackrel{\sim }{Q}^{T}=(Q_{2s},Q_{3s})$, $I$ is a unit matrix, $%
\stackrel{\rightarrow }{\tau }=
\sigma _{x}\widehat{x}+\sigma _{z}\widehat{z}$%
, $\sigma _{x}$ and $\sigma _{z}$ are Pauli matrix, 
$\stackrel{\rightarrow }{%
\Delta K_{(\pi \pi \pi )}}$ $=0$. For 
$\stackrel{\rightarrow }{\kappa }=(\pi
\pi 0),(0\pi \pi ),$and $(\pi 0\pi )$, 
$\stackrel{\rightarrow }{\Delta K_{%
\stackrel{\rightarrow }{\kappa }}}=\Delta K(\sin \psi _{\stackrel{%
\rightarrow }{\kappa }}\widehat{x}+\cos \psi _{\stackrel{\rightarrow }{%
\kappa }}\widehat{z})$, where $\psi _{(\pi \pi 0)}=0$, $\psi _{(0\pi \pi
)}=2\pi /3$, $\psi _{(\pi 0\pi )}=-2\pi /3$.
\bibitem{fn:Q2u}  In our calculation, we will only consider 
square substrate
lattice, which will not generate any $Q_{2u}$ strain. Therefore, we only
need this term for the current study. If we want to find the strain effect
for non-square substrate, we should include the terms with $Q_{2u}$ also.
\bibitem{Drude}  The Drude part is zero, because we deal with insulators
only.
\end{references}
\end{document}